# Stock Portfolio Optimization Using a Deep Learning LSTM Model

Jaydip Sen
*Department of Data Science*
*Praxis Business School*
Kolkata, INDIA
jaydip.sen@acm.org

Abhishek Dutta
*Department of Data Science*
*Praxis Business School*
Kolkata, INDIA
duttaabhishek0601@gmail.com

Sidra Mehtab
*Department of Data Science*
*Praxis Business School*
Kolkata, INDIA
smehtab@acm.org

*Abstract*—Predicting future stock prices and their movement patterns is a complex problem. Hence, building a portfolio of capital assets using the predicted prices to achieve the optimization between its return and risk is an even more difficult task. This work has carried out an analysis of the time series of the historical prices of the top five stocks from the nine different sectors of the Indian stock market from January 1, 2016, to December 31, 2020. Optimum portfolios are built for each of these sectors. For predicting future stock prices, a long-and-short-term memory (LSTM) model is also designed and fine-tuned. After five months of the portfolio construction, the actual and the predicted returns and risks of each portfolio are computed. The predicted and the actual returns of each portfolio are found to be high, indicating the high precision of the LSTM model.

## I. INTRODUCTION

Portfolio Design is a complex optimization problem that is concerned with allocating a given capital to a set of stock or capital assets so that the return and the risk of the investment are optimized. The problem belongs to the NP-hard class of computation for which no algorithm exists that can solve the problem running in polynomial time complexity. The complexity of the problem increases even further if the optimization methods need to have a robust estimation of the returns and risks associated with each constituent stock of the portfolio since accurately estimating the future values of stock prices is also a very challenging problem. In fact, the supporters of the efficient market hypothesis advocate that it is impossible to arrive at a precise predicted value of the future prices of a stock. On the other hand, several researchers and financial analysts have demonstrated how the design of smart algorithms and complex models can help one in predicting future stock prices with high precision. After the seminal proposition of Markowitz on portfolio optimization following the minimum-variance approach, several methods have been suggested by researchers for solving the problem [1]. Several approaches are also proposed in the literature for precisely predicting future stock prices including linear and non-linear regression, *autoregressive integrated moving average* (ARIMA), simple and weighted exponential smoothing, *vector autoregression* (VAR), machine learning, and deep learning.

This work demonstrates a method of constructing optimum and efficient portfolios for nine important sectors of the Indian economy by choosing important stocks from those sectors. After the nine sectors for the study are identified, for each sector, the most critical five stocks are selected based on the report published by the National Stock Exchange (NSE) of India, on May 31, 2021 [2]. The past prices of these 45 stocks as listed in the NSE are extracted from the Yahoo Finance site by the APIs of the Python programming language using their ticker names. Exploiting the features of the past stock prices spanning over five years, portfolios are designed for the nine sectors minimizing the risk involved and optimizing the returns and the risk values. In addition, a deep learning regression model based on LSTM architecture is designed for accurately forecasting future stock prices. After the elapse of five months since portfolio construction, the actual returns yielded by the portfolios and those predicted by the LSTM models are computed. For each sector, the actual and the predicted returns are compared for evaluating the accuracy of the LSTM model, and for having an idea about the actual profitability of investment and risk associated with the sectors under study.

The following are the main contributions of the work. First, based on the past five years' stock prices listed in the NSE for nine critical sectors of the Indian economy, the work presents a step-by-step algorithmic method of designing optimized portfolios for those sectors. These portfolios can be used for making good investment decisions in stocks. serve as an illustrative guide for investors in the Indian stock market. Second, the work proposes an LSTM model for predicting future stock prices. The high prediction accuracy level of the model indicates that it can be used in real-world portfolio construction with a minimum error. Finally, the observations of this study provide a detailed insight into the current return on investments for the nine sectors.

The paper is organized as follows. Section II presents a broad overview of some related works. Section III provides the details of the data used and outlines the methodology followed. Section IV presents the architecture and the top-level design of the LSTM model. Section V presents the results of different portfolios and their actual and predicted returns and risks. The paper is concluded in Section VI.

## II. RELATED WORK

In the literature, several propositions exist for stock price prediction and portfolio optimization. Of late, one of the most popular approaches to stock price prediction is the use of learning-based models and algorithms and artificial intelligence-based systems [3-10]. Integrating text mining in the social web for further improving the prediction accuracy of the learning-based models is also very common [11-13]. Multi-objective optimization approaches for robust portfolio design using metaheuristics and constraint-imposed heuristics have been proposed by many researchers [14-15]. Since the original mean-variance optimization approach proposed by Markowitz has several limitations in practice, purchase limit and cardinality constraint-based approaches are presented by some authors to overcome those shortcomings [16-17]. Several propositions exist in the literature for portfolio optimization using approaches such as fuzzy logic, swarm intelligence, and genetic algorithms, and principal component analysis [18-19]. Generalized autoregressive conditional heteroscedasticity (GARCH) is a widely used approach in estimating risks in a portfolio [20].

In the current work, the minimum-variance and optimum risk approaches for building optimized portfolios have been used for nine important sectors listed in the NSE of India. Based on the historical prices of the stocks for five years from 2016 to 2020, nine portfolios are built. An LSTM model is then designed for predicting the future prices of the stocks in each portfolio. Five months after the portfolio construction, the actual return and the return predicted by the LSTM model are computed and compared.

### III. DATA ACQUISITION AND METHODOLOGY

While the Python programming language has been used in developing the portfolios of stocks, the rich libraries of Tensorflow and Keras frameworks are utilized for designing the LSTM regression model. In the following, the details of the eight-step approach followed in designing the portfolios and the LSTM model, have been discussed.

#### A. Selections of the Sectors and Stocks

First, nine critical sectors of the NSE are chosen for the current study. These sectors are as follows: *pharmaceuticals, infrastructure, realty, media, public sector banks, private sector banks, large-cap, mid-cap, and small-cap*. The NSE publishes reports on a monthly basis identifying the top stocks in each sector and their respective contribution to the computation of the overall sectoral index. The contributions are expressed as weights in percentage. Based on NSE's report published on May 31, 2021, the five most significant stocks for each of the nine sectors are first identified [2]. At the completion of these steps, forty-five stocks are selected for the chosen nine sectors.

#### B. Data Acquisition

For each sector, the historical prices of its five most significant stocks are extracted from the *Yahoo Finance* website using the *DataReader* method available in the *data* sub-module in the *pandas_datareader* module of the Python language. The period for which the stock prices are used for portfolio design spans over five years starting from Jan 1, 2016, and ending on Dec 31, 2020. The raw data extracted from the web have six features: *open, high, low, close, volume*, and *adjusted_close*. Since the current work is based on univariate analysis, the variable *close* for the period Jan 1, 2016, to Dec 31, 2020, is retained for designing the portfolios, and further analysis, while the other variables are ignored.

#### C. Deriving Portfolio Return and Volatility

Based on the historical *close* values of each stock, its return and log return values are computed on daily basis. The daily return and the log return for a stock express the change in the *close* values over successive days and their logarithms, respectively, both computed in percentage values. The daily return and log returns are computed using the library-defined function, *pct_change*, in Python. Next, the daily and yearly volatility values for each stock are computed. The daily volatility of a stock is the standard deviation of its daily returns. The Python function *std* is used for computing the volatility values. The yearly volatility for a stock is derived by multiplying its daily volatility by the square root of 250, assuming 250 working days in a calendar year. From an investor's viewpoint, the yearly volatility of a stock is an indication of the degree of risk associated with it.

#### D. Covariance and Correlation Matrices of Stock Returns

Once the return and volatility values for each stock are computed, the covariance and the correlation matrices of the stocks are derived based on their return values in the training dataset (i.e., from Jan 1, 2016, to Dec 31, 2020). These matrices help one understand the patterns of association among the stocks in a given sector which serve as fundamental inputs to portfolio design. The covariance and the correlation matrices are computed using Python functions *cov* and *corr*, respectively. Risk minimization and optimization is one of the primary optimization objectives in a portfolio design task. In a diversified portfolio that minimizes the risk, the algorithm attempts to allocate funds among the stock that exhibit low or no correlation among them. Identification of such stocks is possible by analyzing their covariance or correlation matrices.

#### E. Estimation of Portfolio Return and Risk

Using the covariance and the correlation matrices for each sector, the first set of portfolios is constructed at this step. At first, the portfolios are designed by assigning equal weights to all five stocks in a given sector. Each stock of a sector is assigned a weight of 0.2 so that their sum is 1. The yearly return and volatility values for the equal-weight portfolio are computed for each of the nine sectors. The computation of the return of a portfolio based on its historical return values are derived using (1), in which, *Exp(Ret)* is the expected return of the portfolio that is consisted of *n* capital assets (i.e., stocks) denoted as $C_1$, $C_2$, …., $C_n$, with the weight $w_i$ assigned to the *i*-th asset.

$$Exp(Ret) = w_1 E(Ret_{C_1}) + w_2 E(Ret_{C_2}) + w_n E(Ret_{C_n}) \quad (1)$$

Using the yearly return and volatility values of the stocks, the yearly return and volatilities of the equal-weight portfolio of each sector are determined. The Python function resample is used with the parameter 'Y' is used for computing the mean yearly returns for each constituent stock in a portfolio. On the other hand, the yearly volatility values for the equal-weight portfolios are derived by multiplying the daily volatility values by a factor of the square root of 250. The equal-weight portfolios provide one with a base level of return and risk associated with the sectors over the training records, and they can be used as benchmarks for evaluating the performance of other portfolios. However, the return and risk estimated using the equal-weight portfolios serve as very poor estimators of future returns and risks. Hence, more precise estimations of the future return and risks are needed. This necessitates the design of the minimum-risk and the optimum-risk portfolios.

## F. Designing Minimum-Risk Portfolios

To improve on the equal-weight portfolios, at this step, minimum-risk portfolios are designed for the nine sectors. The minimum-risk portfolios have the minimum values for their variances. The variance of a given portfolio, *Variance(P)*, is dependent on the variances of its constituent stocks and the covariances among each pair as given by (2).

$$Variance(P) = \sum_{i=1}^{n} w_i s_i^2 + 2 * \sum_{i,j} w_i * w_j * Covar(i,j) \quad (2)$$

In (2), the volatility measured by the standard deviation of the stock i is represented by si, the weight allocated to it is denoted by $w_i$, and the covariance among the returns of stock *i* and stock *j* is denoted as *Covar(i, j)*. Since each portfolio consists of five stocks, the computation of its variance involves 15 terms. Five of the terms are involved in the weighted sum of the variances of the individual stocks, while the remaining ten terms reflect the covariances between each pair. The minimum risk portfolio is that portfolio that identifies the combination of $w_i$'s that leads to the minimum value of the volatility of the portfolio.

For identifying the minimum-risk portfolio for each sector, the *efficient frontier* (EF) of several portfolios for that sector is first visualized. The EF of a set of portfolios in two-dimensional space exhibits a plot in which the return and volatility are depicted along the *y*-axis and the *x*-axis, respectively. The EF frontier consists of those points (i.e., portfolios), that yield the highest return for a given value of volatility, or the lowest volatility for a given value of the return. The point that occupies the leftmost position on the EF signifies the portfolio with the lowest volatility, hence it represents the minimum-risk portfolio. For constructing the contour of the efficient frontier, weights are assigned randomly to the five stocks in a portfolio in a loop, and the loop is iterated 10,100 times in a Python program. The program outputs 10, 000 points each of which corresponds to a portfolio with a specified return and risk values. The points on the EF are those which yield the lowest volatility for a given return level or the highest return value for a given volatility. The leftmost point on the EF among all the generated points represents the minimum-risk portfolio.

## G. Designing Optimum-Risk Portfolios

The investors in the stock market rarely follow the strategy of risk minimization as proposed by the minimum-risk portfolio due to their low returns. Most often, the investors are ready to undertake higher risks if the associated returns are even higher. For optimizing the risk and return in a portfolio, and thereby designing an optimum-risk portfolio, a metric called Sharpe Ratio is used, which is given by (3). For computing the optimum risk portfolio, the metric Sharpe Ratio of a portfolio is used. The Sharpe Ratio of a portfolio is given by (3).

$$Sharpe\ Ratio = \frac{Ret_{curr} - Ret_{risk\_free}}{Ris_{curr}} \quad (3)$$

In (3), $Ret_{curr}$, $Ret_{risk\_free}$, and $Ris_{curr}$, denote respectively, the current portfolio return, the risk-free portfolio return, and the current portfolio risk (measured by its yearly standard deviation). The portfolio with a risk value of 1% is assumed to be a risk-free one. The optimum-risk portfolio maximizes the value of the Sharpe Ratio. A substantially high return with a small increase in risk is achieved by the optimum-risk portfolio when compared with the minimum-risk portfolio. Among all the candidate portfolios on the EF, the portfolio with the highest Sharpe Ratio is identified using the *idmax* function in Python.

## H. Predicted and Actual Return and Risks of Portfolios

Using the training dataset from Jan 1, 2016, to Dec 31, 2020, a couple of portfolios are built for the sectors – a minimum risk portfolio and an optimal risk portfolio. On January 1, 2021, a fictitious investor is created who invests an amount of Indian Rupees (INR) of 100000 for each sector based on the recommendation of the optimal risk portfolio structure for the corresponding sector. Note that the amount of INR 100000 is just for illustrative purposes only. The analysis will not be affected either by the currency or by the amount. To compute the future values of the stock prices and hence to predict the future value of the portfolio, a regression model is designed using the LSTM deep learning architecture. On May 31, 2021, using the LSTM model, the stock prices for June 1, 2021, are predicted (i.e., a forecast horizon of one week is used). Based on the predicted stock values, the predicted rate of return for each portfolio is determined. And finally, on June 1, 2021, when the actual prices of the stocks are known, the actual rate of return is determined. The predicted and actual rates of return for the portfolios are compared to evaluate the profitability of the portfolios and the prediction accuracy of the LSTM model.

## IV. THE LSTM MODEL ARCHITECTURE

As explained in Section III, the stock prices are predicted with a forecast horizon of one day, using an LSTM deep learning model. This section presents the details of the architecture and the choice of various parameters in the model design. First, a very brief discussion on the fundamentals of LSTM networks and the effectiveness of these networks in interpreting sequential data is discussed is presented. Then the design details of the proposed model are presented.

LSTM is an extended and advanced, recurrent neural network (RNN) with a high capability of interpreting and predicting future values of sequential data like time series of stock prices or text [21]. LSTM networks are able to maintain their state information in some specially designed memory cells or gates. The networks carry out an aggregation operation on the historical state stored in the forget gates with the current state information to compute the future state information. The information available at the current time slot is received at the input gates. Using the results of aggregation at the forget gates and the input gates, the networks predict the next value of the target variable. The predicted value is available at the output gates [21].

For predicting the future values of stock prices, an LSTM model is designed and fine-tuned. The schematic design of the model is exhibited in Fig. 1. The model uses daily close prices of the stock of the past 50 days as the input. The input data of 50 days with a single feature (i.e., *close* values) has a shape (50, 1). The input layer receives and forwards the data to the first LSTM layer, which is composed of 256 nodes. The LSTM layer yields a shape of (50, 256) at its output. This implies that 256 features are extracted by each node of the LSM layer from every record in the input data. A dropout layer is used after the first LSTM layer that randomly switches off the output of thirty percent of the nodes in the LSTM to avoid model overfitting. Another LSTM layer with the same architecture as the previous one receives the output from the first and applies a dropout rate of thirty percent. A dense layer with 256 nodes receives the output from the second LSTM layer. A single node at the output of the dense layer yields the predicted value of the *close* price. The forecast horizon can be adjusted to different values by adjusting a tunable parameter. A forecast horizon of one day is used so that a prediction for the next day is made. A batch size of 64 and 100 epochs are used for training and validating the model. Except for the output layer, the activation function *rectified linear unit* (ReLU) is used. At the output layer, the *sigmoid* activation function is used. The training and validation accuracies and losses are measured using the *mean absolute error* (MAE) function and Huber Loss, respectively. The optimum epoch no and batch size are found using grid-search.

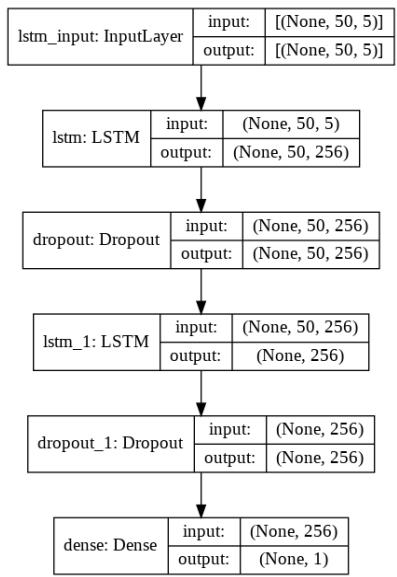

Fig. 1. The schematic diagram of the LSTM model

It may be mentioned that the hyperparameter values used in the network are all determined using the grid-search method. The Huber loss function is used because of its adaptability in combining the characteristics of mean MSE) and MAE [21].

## V. RESULTS

In this section, the performance results of the nine portfolios are presented, and a detailed analysis is carried out on the results. The stocks are chosen from the following sectors: (i) *pharmaceuticals*, (ii) *infrastructure*, (iii) *realty*, (iv) *media*, (v) *public sector banks*, (vi) private sector banks, (vii) *large-cap*, (viii) *mid-cap*, and (ix) *small-cap*. In implementing the portfolio models and the LSTM model, Python language (version 3.7.4) has been used. Various useful modules and libraries of the TensorFlow 2.3.0 framework and Keras 2.4.5 have also been utilized. The GPU of the Google Colab environment is used for building the models and their testing [22].

### A. Pharmaceuticals Sector Stocks

The five critical stocks of the pharma sector and their weights used in computing the pharma sector index as reported by the NSE on May 31, 2021, are mentioned in the following: Sun Pharmaceuticals Ind. (SNP): 20.34, Dr. Reddy's Lab (DRL): 18.17), Divi's Lab (DVL): 15.07, Cipla (CPL): 13.60, and Lupin (LUP): 8.27 [2].

TABLE I PHARMA SEC PORTFOLIOS: WEIGHT ALLOCATION

| Stocks | Min Risk | Opt Risk |
|---|---|---|
| SNP | 0.1163 | 0.0202 |
| DRL | 0.3030 | 0.1907 |
| DVL | 0.1250 | 0.6073 |
| CPL | 0.3277 | 0.1378 |
| LUP | 0.1281 | 0.0441 |
| Annual Return (%) | 16.24 | 36.92 |
| Annual Risk (%) | 22.19 | 27.08 |

Table I shows the weights allocated by the min risk and the opt risk portfolios to the stocks of the pharma sector. Table II presents the actual and the predicted return of the optimum portfolio over five months (i.e., from Jan 1, 2021, to Jun 1, 2021) as computed on June 1, 2021. Fig. 2 shows the efficient frontier, the min risk, and the opt risk portfolios of the pharma sector. Fig 3 depicts the actual prices and their predicted values for SNP, the leading stock of this sector.

TABLE II ACTUAL AND PREDICTED RETURN OF PHARMA PORTFOLIO

| Stock | Date: Jan 1, 2021 | | | Date: Jun 1, 2021 | | | |
|---|---|---|---|---|---|---|---|
| | *Amt Invstd* | *Act Price* | *No of Stocks* | *Act Price* | *Act Val* | *Pred Price* | *Pred Val* |
| SNP | 2020 | 596 | 3.39 | 671 | 2275 | 669 | 2268 |
| DRL | 19070 | 5241 | 3.64 | 5317 | 19354 | 5295 | 19274 |
| DVL | 60730 | 3849 | 15.78 | 4220 | 66592 | 4232 | 66781 |
| CPL | 13780 | 827 | 16.66 | 946 | 15760 | 956 | 15926 |
| LUP | 4400 | 1001 | 4.40 | 1209 | 5320 | 1195 | 5258 |
| Total | 100000 | | | | 109301 | | 109507 |
| ROI (%) | **Actual: 9.30** | | | **Predicted:** 9.51 | | | |

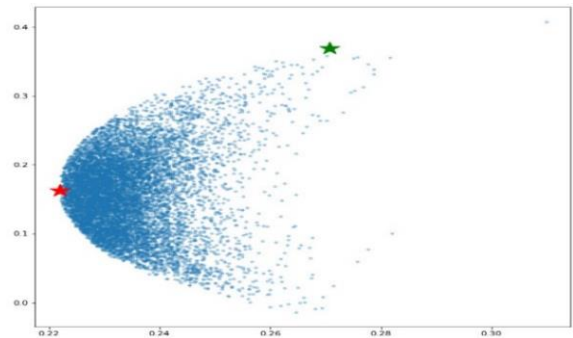

Fig. 2. Min risk and the opt risk portfolios, marked by the red and the green star, respectively, for the pharma sector stocks built on Jan 1, 2021. The *x*- and the *y*-axis plot the risk and return, respectively.

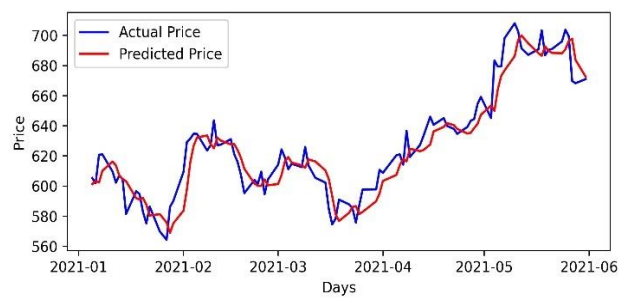

Fig. 3. Sun Pharma (SNP) stock: actual vs. predicted prices by the LSTM model (Period: Jan 1, 2021, to Jun 1, 2021)

266

TABLE III  INFRASTRUCTURE SEC PORTFOLIOS: WEIGHT ALLOCATION

| Stocks | Min Risk | Opt Risk |
|---|---|---|
| RIL | 0.1733 | 0.7613 |
| LNT | 0.1523 | 0.0194 |
| BAT | 0.1141 | 0.1694 |
| UTC | 0.1254 | 0.0235 |
| PGC | 0.4349 | 0.0263 |
| Annual Return (%) | 13.72 | 35.26 |
| Annual Risk (%) | 18.96 | 26.10 |

*B. Infrastructure Sector Stocks*

The five critical stocks of the infrastructure sector and their weights are as follows: Reliance Industries (RIL): 20.02, Larsen & Toubro (LNT): 12.91, Bharti Airtel (BAT): 9.41, Ultra Tech Cement (UTC): 5.64, Power Grid Corporation of India (PGC): 4.21 [2]. Tables III and IV present the infrastructure sector's results. Fig. 4 exhibits the efficient frontier, while Fig. 5 shows the actual prices and their predicted values for RIL, the leading stock of this sector.

TABLE IV  ACTUAL AND PREDICTED RETURN OF INFRA PORTFOLIO

| Stock | Date: Jan 1, 2021 | | | Date: Jun 1, 2021 | | | |
|---|---|---|---|---|---|---|---|
|  | Amt Invstd | Act Price | No of Stocks | Act Price | Act Val | Pred Price | Pred Val |
| RIL | 76130 | 1988 | 38.29 | 2169 | 83051 | 2080 | 79643 |
| LNT | 1950 | 1297 | 1.50 | 1475 | 2213 | 1451 | 2177 |
| BAT | 16940 | 515 | 32.89 | 533 | 17530 | 526 | 17300 |
| UTC | 2350 | 5291 | 0.44 | 6601 | 2904 | 6602 | 2905 |
| PGC | 2630 | 190 | 13.84 | 224 | 3100 | 222 | 3072 |
| Total | 100000 | | | | 108798 | | 105097 |
| ROI (%) | Actual: 8.80    Predicted: 5.10 | | | | | | |

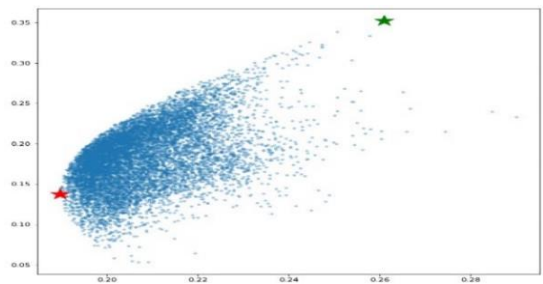

Fig. 4. Min. risk and opt risk portfolios, marked by the red and the green star, respectively, for the infrastructure sector stocks built on Jan 1, 2021. The *x*- and the *y*-axis plot risk and return, respectively.

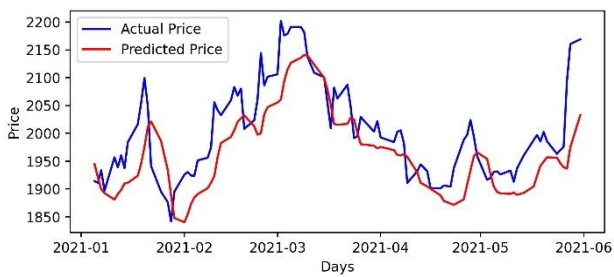

Fig. 5. Reliance Ind. (RIL) stock: actual vs. predicted prices by the LSTM model (Period: Jan 1, 2021, to Jun 1, 2021)

*C. Realty Sector Stocks*

The five stocks that contribute most significantly to the realty sector index and the corresponding weights (in percent) are as follows. DLF (DLF): 26.70, Godrej Properties (GPR): 23.92, Oberoi Realty (OBR): 11.72, Phoenix Mills (PHM):10.98, and Prestige Estates Projects (PRE): 6.85 [2]. Tables V and VI present the results of the realty sector portfolio. Fig. 6 shows the efficient frontier, while Fig. 7 shows the actual and forecasted prices by the LSTM model for the leading stock of the realty sector DLF.

TABLE V  REALTY SEC PORTFOLIOS: WEIGHT ALLOCATION

| Stocks | Min Risk | Opt Risk |
|---|---|---|
| DLF | 0.0983 | 0.0646 |
| GRP | 0.2398 | 0.6309 |
| OBR | 0.2259 | 0.0379 |
| PHM | 0.3382 | 0.2337 |
| PRE | 0.0978 | 0.0330 |
| Annual Return (%) | 31.61 | 43.92 |
| Annual Risk (%) | 27.39 | 31.86 |

TABLE VI  ACTUAL AND PREDICTED RETURN OF REALTY PORTFOLIO

| Stock | Date: Jan 1, 2021 | | | Date: Jun 1, 2021 | | | |
|---|---|---|---|---|---|---|---|
|  | Amt Invstd | Act Price | No of Stocks | Act Price | Act Val | Pred Price | Pred Val |
| DLF | 6460 | 238 | 27.14 | 287 | 7789 | 281 | 7626 |
| GRP | 63090 | 1427 | 44.21 | 1361 | 60170 | 1368 | 60479 |
| OBR | 3780 | 590 | 6.41 | 605 | 3878 | 594 | 3808 |
| PHM | 23370 | 785 | 29.77 | 791 | 23548 | 782 | 23280 |
| PRE | 3300 | 267 | 12.36 | 272 | 3362 | 274 | 3387 |
| Total | 100000 | | | | 98747 | | 98580 |
| ROI (%) | Actual: -1.25    Predicted: -1.42 | | | | | | |

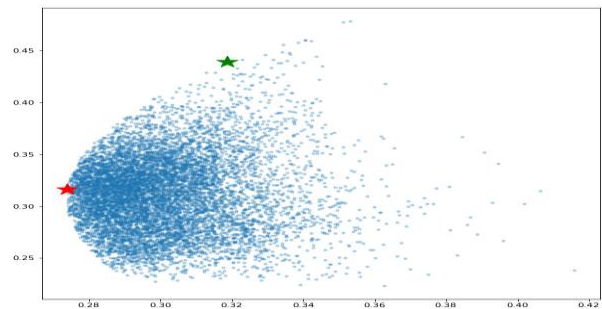

Fig. 6. Min. risk and opt risk portfolios, marked by the red and the green star, respectively, for the realty sector stocks built on Jan 1, 2021. The *x*- and *y*-axis plot risk and return, respectively.

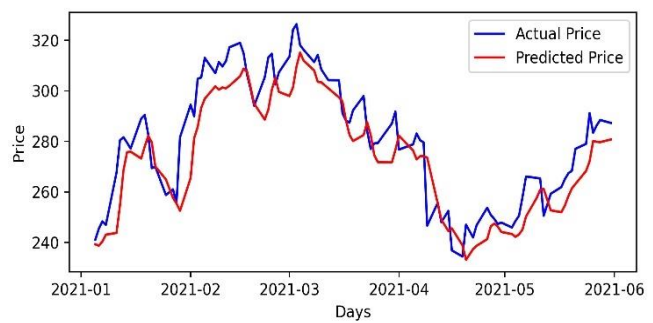

Fig. 7. DLF stock: actual vs. predicted prices by the LSTM model (Period: Jan 1, 2021, to Jun 1, 2021)

TABLE VII  MEDIA SEC PORTFOLIOS: WEIGHT ALLOCATION

| Stocks | Min Risk | Opt Risk |
|---|---|---|
| ZEE | 0.1788 | 0.0120 |
| PVR | 0.3981 | 0.4020 |
| STN | 0.2373 | 0.4282 |
| TVB | 0.1388 | 0.1548 |
| DTI | 0.0471 | 0.0029 |
| Annual Return (%) | 2.43 | 8.17 |
| Annual Risk (%) | 29.03 | 30.58 |

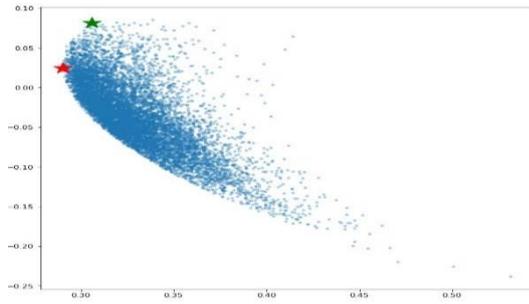

Fig. 8. Min risk and opt risk portfolios, marked by the red and the green star, respectively, for the media sector stocks built on Jan 1, 2021. The *x*- and the *y*-axis plot the risk and the return, respectively.

### D. Media Sector Stocks

The most critical stocks of the media sector and the respective contributions (in percent) of those stocks in the computation of the media sector index are the following. Zee Entertainment Enterprises (ZEE): 27.16, PVR (PVR): 16.44, Sun TV Network (STN): 15.97, TV18 Broadcast (TVB): 12.51, Dish TV India (DTI): 7.93 [2]. Tables VII and VIII present the results of the media sector portfolio. Fig. 8 and Fig 9 respectively exhibit the efficient frontier of the media sector portfolio, and the actual prices, and the corresponding forecasted prices by the LSTM model for the ZEE stock, the most significant stock of this sector.

TABLE VIII  ACTUAL AND PREDICTED RETURN OF MEDIA PORTFOLIO

| Stock | Date: Jan 1, 2021 | | | Date: Jun 1, 2021 | | | |
|---|---|---|---|---|---|---|---|
| | *Amt Invstd* | *Act Price* | *No of Stocks* | *Act Price* | *Act Val* | *Pred Price* | *Pred Val* |
| ZEE | 1200 | 225 | 5.33 | 213 | 1135 | 215 | 1146 |
| PVR | 40200 | 1340 | 30.00 | 1307 | 39210 | 1312 | 39360 |
| STN | 42820 | 478 | 89.58 | 526 | 47119 | 538 | 48194 |
| TVB | 15490 | 31 | 499.68 | 41 | 20487 | 42 | 20987 |
| DTI | 290 | 13 | 22.31 | 15 | 335 | 16 | 357 |
| Total | 100000 | | | | 108286 | | 110044 |
| ROI (%) | Actual: 8.29 | | | Predicted: 10.04 | | | |

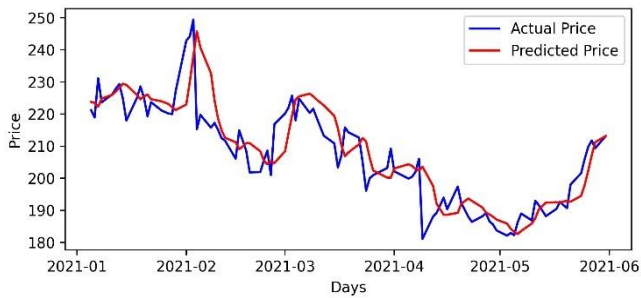

Fig. 9. Zee Entertainment (ZEE) stock: actual vs. the predicted prices by the LSTM model (Period: January 1, 2021, to June 1, 2021)

### E. Public Sector Banks Stocks

The stocks that have the most impactful contributions to the overall sectoral index of the NIFTY public sector banks and their respective contributions (in percent) are as follows. State Bank of India (SBI): 28.47, Bank of Baroda (BOB): 18.25, Punjab National Bank (PNB): 15.31, Canara Bank (CNB): 15.24, and Bank of India (BOI): 5.23 [2]. Tables IX and X exhibit the results of the portfolios of this sector. Fig.10 and Fig. 11 show the efficient frontier of the sector and the actual prices and their predicted values for SBI, the leading stock of this sector.

TABLE IX  PSU BANKS PORTFOLIOS: WEIGHT ALLOCATION

| Stocks | Min Risk | Opt Risk |
|---|---|---|
| SBI | 0.6281 | 0.7312 |
| BOB | 0.0017 | 0.0414 |
| PNB | 0.0582 | 0.0456 |
| CNB | 0.0376 | 0.1401 |
| BOI | 0.2744 | 0.0418 |
| Annual Return (%) | -2.21 | -0.94 |
| Annual Risk (%) | 37.25 | 37.46 |

TABLE X  ACT. AND PRED. RET. OF PUBLIC SECTOR BANK PORTFOLIO

| Stock | Date: Jan 1, 2021 | | | Date: Jun 1, 2021 | | | |
|---|---|---|---|---|---|---|---|
| | *Amt Invstd* | *Act Price* | *No of Stocks* | *Act Price* | *Act Val* | *Pred Price* | *Pred Val* |
| SBI | 73120 | 279 | 262.08 | 433 | 113480 | 418 | 109549 |
| BOB | 4140 | 65 | 63.69 | 79 | 5032 | 81 | 5159 |
| PNB | 4560 | 35 | 130.29 | 42 | 5472 | 42 | 5472 |
| CNB | 14010 | 133 | 105.34 | 159 | 16749 | 161 | 16959 |
| BOI | 4170 | 50 | 83.40 | 76 | 6338 | 76 | 6338 |
| Total | 100000 | | | | 147071 | | 143478 |
| ROI (%) | Actual: 47.07 | | | Predicted: 43.48 | | | |

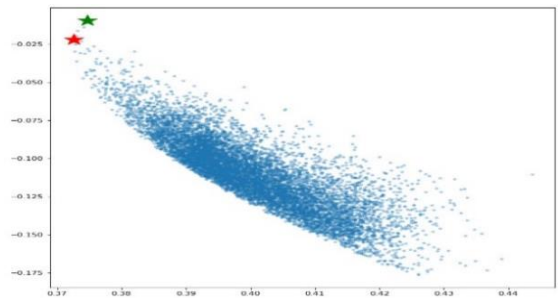

Fig. 10. Min risk and opt risk portfolios marked by the red and the green star, respectively, for the PSU banks' stocks built on Jan 1, 2021. The x- and the y-axis plot the risk and the return, respectively.

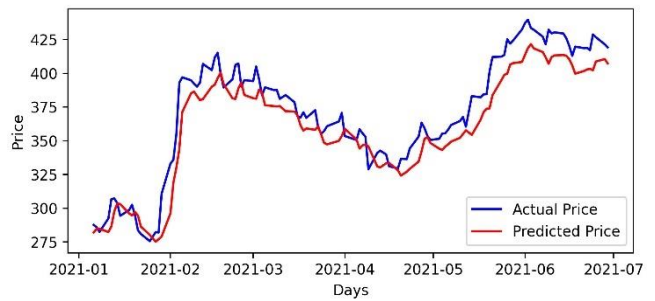

Fig. 11. State Bank of India (SBI) stock: actual vs. predicted prices by the LSTM model (Period: January 1, 2021, to June 1, 2021)

TABLE XI  PVT SEC BANKS PORTFOLIOS: WEIGHT ALLOCATION

| Stocks | Min Risk | Opt Risk |
|---|---|---|
| HDB | 0.5639 | 0.6187 |
| ICB | 0.0708 | 0.0344 |
| AXB | 0.0623 | 0.0010 |
| KMB | 0.2933 | 0.3351 |
| IIB | 0.0096 | 0.0108 |
| Annual Return (%) | 25.25 | 26.36 |
| Annual Risk (%) | 22.90 | 22.92 |

### F. Private Sector Banks Stocks

The stocks with the most critical influence on the sectoral index of the NIFTY private sector banks and their respective percentage-wise contributions are as follows. HDFC Bank (HDB): 25,52, ICICI Bank (ICB): 23.24, Axis Bank (AXB): 15.00, Kotak Mahindra Bank (KMB): 13.44, and IndusInd

Bank (IIB): 9.46 [2]. Tables XI and XII present the results of the portfolio of the private sector banks. Fig. 12 and Fig. 13, respectively, show the efficient frontier of the sector and the actual prices and their predicted values for the most critical stock of this sector, HDFC Bank.

TABLE XII  ACT. AND PRED. RETURN OF PVT. SEC. BANK PORTFOLIO

| Stock | Date: Jan 1, 2021 | | | Date: Jun 1, 2021 | | | |
|---|---|---|---|---|---|---|---|
| | Amt Invstd | Act Price | No of Stocks | Act Price | Act Val | Pred Price | Pred Val |
| HDB | 61870 | 1425 | 43.42 | 1512 | 65647 | 1508 | 65474 |
| ICB | 3440 | 528 | 6.52 | 650 | 4235 | 642 | 4183 |
| AXB | 100 | 624 | 0.16 | 745 | 119 | 739 | 118 |
| KMB | 33510 | 1994 | 16.81 | 1797 | 30199 | 1782 | 29947 |
| IIB | 1080 | 900 | 1.20 | 1010 | 1213 | 1012 | 1214 |
| Total | 100000 | | | | 101413 | | 100936 |
| ROI (%) | Actual: 1.41 | | | Predicted: 0.94 | | | |

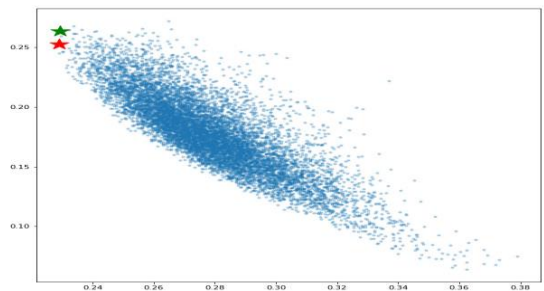

Fig. 12. Min risk and opt risk portfolios marked by the red and the green star, respectively, for the private banks' stocks built on Jan 1, 2021. The *x*- and the *y*-axis plot the risk and the return, respectively.

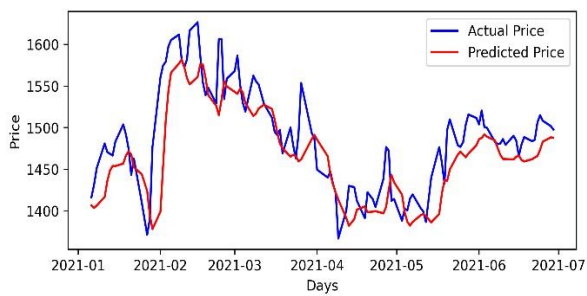

Fig. 13. HDFC Bank (HDB) stock: actual vs. the predicted prices by the LSTM model (Period: January 1, 2021, to June 1, 2021)

TABLE XIII  LARGE-CAP SEC PORTFOLIOS: WEIGHT ALLOCATION

| Stocks | Min Risk | Opt Risk |
|---|---|---|
| RIL | 0.1970 | 0.4648 |
| HDB | 0.4404 | 0.1961 |
| IFY | 0.3107 | 0.3088 |
| HDF | 0.0455 | 0.0179 |
| ICB | 0.0134 | 0.0124 |
| Annual Return (%) | 28.68 | 32.81 |
| Annual Risk (%) | 19.70 | 21.40 |

### G. Large-Cap Sector Stocks

The stocks that have a critical influence on the sectoral index of the large-cap sector and their respective contributions are as follows. Reliance Industries (RIL): 10.36, HDFC Bank (HDB): 9.79, Infosys (IFY): 7.66, Housing Development Finance Corporation (HDF): 6.82, and ICICI Bank (ICB): 6.80 [2]. Tables XIII and XIV present the results of the portfolios of the sector. Fig. 14 depicts the efficient frontier of the portfolios of this sector.

In Fig. 15, the actual and predicted prices for the stock Infosys are depicted, as the same plots for the first two stocks of this sector, RIL, and HDB, have been shown already under the analysis of the infrastructure sector and the private sector banks.

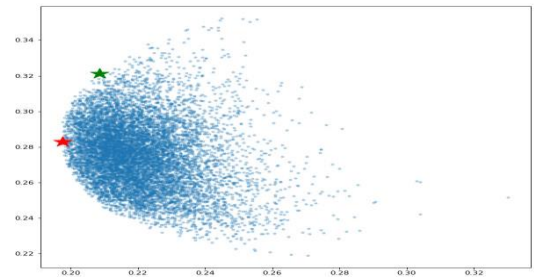

Fig. 14. Min risk and opt risk portfolios marked by the red and the green star, respectively, for the large-cap sector stocks built on Jan 1, 2021. The x- and the y-axis plot the risk and the return, respectively.

TABLE XIV  ACT. AND PRED. RETURN OF LARGE-CAP PORTFOLIO

| Stock | Date: Jan 1, 2021 | | | Date: Jun 1, 2021 | | | |
|---|---|---|---|---|---|---|---|
| | Amt Invstd | Act Price | No of Stocks | Act Price | Act Val | Pred Price | Pred Val |
| RIL | 46480 | 1988 | 23.38 | 2169 | 50711 | 2080 | 48630 |
| HDB | 19610 | 1425 | 13.76 | 1512 | 20805 | 1508 | 20750 |
| IFY | 30880 | 1260 | 24.51 | 1387 | 33995 | 1414 | 34657 |
| HDF | 1790 | 2569 | 0.70 | 2581 | 1807 | 2524 | 1767 |
| ICB | 1240 | 528 | 2.35 | 650 | 1528 | 642 | 1509 |
| Total | 100000 | | | | 108846 | | 107313 |
| ROI (%) | Actual: 8.85 | | | Predicted: 7.31 | | | |

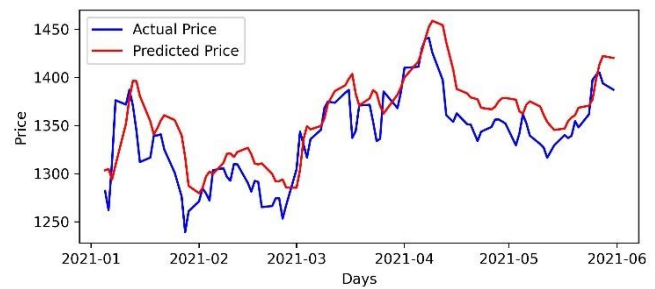

Fig. 15. Infosys (IFY) stock price: actual vs. the predicted values by the LSTM model (Period: January 1, 2021, to June 1, 2021)

TABLE XV  MID-CAP SEC PORTFOLIOS: WEIGHT ALLOCATION

| Stocks | Min Risk | Opt Risk |
|---|---|---|
| STF | 0.0354 | 0.0055 |
| VLT | 0.4560 | 0.8500 |
| CIF | 0.0588 | 0.0497 |
| ASF | 0.2628 | 0.0716 |
| MFS | 0.1871 | 0.0231 |
| Annual Return (%) | 20.10 | 29.25 |
| Annual Risk (%) | 27.14 | 30.22 |

TABLE XVI  ACTUAL AND PREDICTED RETURN OF MID-CAP PORTFOLIO

| Stock | Date: Jan 1, 2021 | | | Date: Jun 1, 2021 | | | |
|---|---|---|---|---|---|---|---|
| | Amt Invstd | Act Price | No of Stocks | Act Price | Act Val | Pred Price | Pred Val |
| STF | 550 | 1071 | 0.51 | 1411 | 725 | 1408 | 723 |
| VLT | 85000 | 831 | 102.29 | 1013 | 103616 | 1011 | 103412 |
| CIF | 4970 | 411 | 12.09 | 546 | 6602 | 543 | 6566 |
| ASF | 7160 | 875 | 8.18 | 989 | 8093 | 983 | 8044 |
| MFS | 2320 | 691 | 3.36 | 936 | 3143 | 911 | 3059 |
| Total | 100000 | | | | 122179 | | 121803 |
| ROI (%) | Actual: 22.18 | | | Predicted: 21.80 | | | |

## H. Mid-Cap Sector Stocks

The five critical stocks form the mid-cap sector and their percentage weights used in the sectoral index computation in NSE are: Shriram Transport Finance Company (STF): 4.21, Voltas (VLT): 3.67, Cholamandalam Investment and Finance (CIF): 3.40, AU Small Finance Bank (ASF): 3.30, and Max Financial Services (MFS): 3.09 [2]. Tables XV and XVI present the results of the portfolios of the mid-cap sector. Fig. 16 depicts the efficient frontier of the portfolio. In Fig. 17, the actual prices of the STF stock and their corresponding predicted values by the LSTM model are depicted. STF is the leading stock of the mid-cap sector.

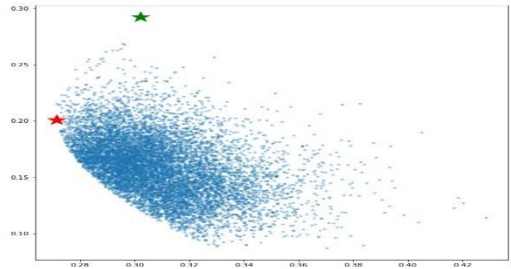

Fig. 16. Min risk and opt risk portfolios marked by the red and the green star, respectively, for the mid-cap sector stocks built on Jan 1, 2021. The *x*- and *y*-axis plot the risk and the return, respectively.

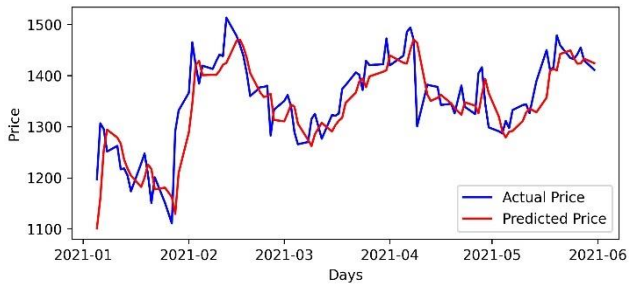

Fig. 17. STF stock price: actual vs. predicted prices by the LSTM model (Period: January 1, 2021, to June 1, 2021)

TABLE XVII  SMALL-CAP SECTOR STOCKS PORTFOLIOS

| Stocks | Min Risk | Opt Risk |
|---|---|---|
| CDS | 0.1815 | 0.0022 |
| KJC | 0.3548 | 0.0687 |
| AAT | 0.1646 | 0.9194 |
| MCE | 0.1908 | 0.0071 |
| IDF | 0.1083 | 0.0027 |
| Annual Return (%) | 24.73 | 63.72 |
| Annual Risk (%) | 24.73 | 35.79 |

TABLE XVIII  ACT. AND PRED. RETURN OF SMALL-CAP PORTFOLIO

| Stock | Date: Jan 1, 2021 | | | Date: Jun 1, 2021 | | | |
|---|---|---|---|---|---|---|---|
|  | Amt Invstd | Act Price | No of Stocks | Act Price | Act Val | Pred Price | Pred Val |
| CDS | 2220 | 532 | 4.17 | 965 | 4027 | 951 | 3968 |
| KJC | 6870 | 709 | 9.70 | 966 | 9360 | 967 | 9370 |
| AAT | 91940 | 860 | 106.91 | 1297 | 138658 | 1344 | 143683 |
| MCE | 710 | 1748 | 0.41 | 1556 | 632 | 1555 | 632 |
| IDF | 270 | 37 | 7.30 | 58 | 423 | 59 | 431 |
| Total | 100000 |  |  |  | 153100 |  | 158084 |
| ROI (%) | Actual: 53.10 | | | Predicted: 58.08 | | | |

## I. Small-Cap Sector Stocks

The most five most important stocks of this sector are Central Depository Services (CDS): 4.17, Kajaria Ceramics (KJC): 4.03, APL Apollo Tubes (AAT): 4.01, Multi Commodity Exchange (MCE): 3.98, and IDFC (IDF): 3.85 [2]. The efficient frontier of the portfolios, and the actual and the predicted prices of the leading stock of this sector, Central Depository Services are depicted in Fig 18 and Fig 19, respectively. Tables XVII and XVIII depict the results of the portfolios of the small-cap sector.

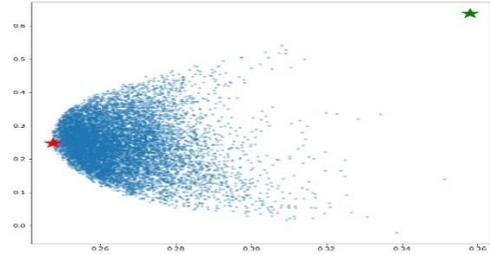

Fig. 18. Min risk and opt risk portfolios marked by the red and the green star, respectively, for the small-cap sector stocks built on Jan 1, 2021. The *x*- and the *y*-axis plot the risk and the return, respectively.

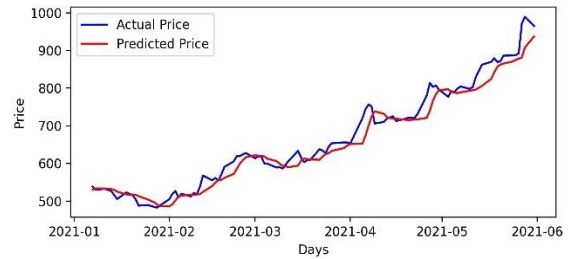

Fig. 19. Central Depository Services (CDS) stock price: actual vs. the predicted values by the LSTM model (Period: Jan 1, 2021, to Jun 1, 2021)

## J. Summary of the Results

The summary of the results of the work is presented in Table XIX, which depicts the actual and the predicted returns of all nine portfolios. It is observed that while the small-cap sector portfolio has yielded the highest rate of return over the five months (i.e., Jan 1, 2021, to Jun 1, 2021), the only sector that yielded a negative return is the realty sector. It is also seen that the accuracy of the LSTM model is high as the predicted returns are quite close to the actual return values.

TABLE XIX  THE SUMMARY OF THE RESULTS

| Portfolio | Act Return (%) | Pred Return (%) |
|---|---|---|
| Pharmaceuticals | 9.30 | 9.51 |
| Infrastructure | 8.80 | 5.10 |
| Realty | -1.25 | -1.42 |
| Media | 8.29 | 10.04 |
| PSU Banks | 47.07 | 43.48 |
| Pvt. Banks | 1.41 | 0.94 |
| Large-Cap | 8.85 | 7.31 |
| Mid-Cap | 22.18 | 21.80 |
| Small-Cap | 53.10 | 58.08 |

## VI. CONCLUSION

The paper has presented nine optimized portfolios for nine important sectors listed in the NSE of India, based on the historical stock prices from Jan 1, 2010, to Dec 31, 202. An LSTM model is also designed for predicting future stock prices. After a hold-out period of five months, the actual and the predicted returns of each portfolio are computed, and their values are compared for evaluating the accuracy of the predictive model. The model is found to be highly accurate in predicting stock prices over a short time horizon. In future work, the performances of the optimum portfolios will be compared with those of other approaches like eigen portfolios.